\documentclass[letterpaper,twoside,10pt]{article}
\usepackage{amssymb}
\usepackage{amsmath2000} 
\usepackage{latexsym}
\usepackage{verbatim}
\usepackage{epsfig}
\usepackage{fancyheadings}
\usepackage{pstricks,pst-node}
\usepackage{rotating,graphicx}
\newlength{\figurewidth}
\newlength{\enviropost}
\newcommand{\be}{\begin{equation}}
\newcommand{\ee}{\end{equation}}
\newcommand{\ble}[1]{\begin{equation} \label{#1}}
\newcommand{\bae}{\begin{eqnarray}}
\newcommand{\eae}{\end{eqnarray}}
\newcommand{\fle}[2]%
{\vspace{1.5ex}
\be
\label{#1}
\mbox{%
\setlength{\fboxsep}{3ex}%
\framebox{$\dss #2 $}}
\ee} 

\newcommand{\nn}{\nonumber}
\newcommand{\ff}{\nn \\}

\newtheorem{state}{S$\! \!$}
\newtheorem{defin}{D$\! \!$}
\newtheorem{exatitle}{Example}
{\noindent \textsc{Proof}:\ }%
{\hfill $\blacksquare$  \vspace{\enviropost} \\}
{\begin{exatitle} \label{#2} #1 \end{exatitle}}%
{\hfill $\Box$ \vspace{\enviropost} \\}
{\vspace{3mm} \noindent {\bf Aside:} \small}%
{\hfill \rule[-3mm]{0mm}{0mm}$\Diamond$\\}
\newenvironment{statement}[1]%
{%
\vspace{3mm}  
\begin{center}
\begin{minipage}{.8\textwidth} 
\begin{state} 
\label{#1}%
}%
{%
\end{state}
\end{minipage}
\end{center}
\vspace{3mm}
}
\newenvironment{definition}[1]%
{%
\vspace{3mm}  
\begin{center}
\begin{minipage}{.8\textwidth} 
\begin{defin} 
\label{#1}%
}%
{%
\end{defin}
\end{minipage}
\end{center}
\vspace{3mm}
}

\newcommand{\dss}{\displaystyle}

\newcommand{\ket}[1]{| #1 \rangle}

\newcommand{\ipq}[2]{\left\langle #1 \, | \, #2\right\rangle}
\newcommand{\sqrti}{\frac{1}{\sqrt{2}}}

\newcommand{\eg}{\hbox{\em e.g.{}}}

\newcommand{\ie}{\hbox{\em i.e.{}}}

\newcommand{\rhs}{\hbox{r.h.s.{}}}

\newcommand{\capitem}[1]{\caption{\textsf{#1}}}

\newcommand{\calO}{\mathcal{O}}

        \def\ba{\nopagebreak[3]\begin{eqnarray}}
        \def\ea{\end{eqnarray}}

\newcommand{\teta}{\rlap{\lower2ex\hbox{$\,\tilde{}$}}\eta{}}

\newcommand{\partp}{\text{top}}
\newcommand{\partm}{\text{bottom}}
\newcommand{\papertitle}{%
On the Geometrical Character of Gravitation%
}
\newcommand{\paperauthor}{%
Chryssomalis Chryssomalakos and Daniel Sudarsky%
}
\begin{document}
\begin{titlepage}
\begin{flushright}
\textsf{March 26, 2002}
\\
\textsf{}
\\
\textsf{}
\\[3.5cm]
\end{flushright}
\renewcommand{\thefootnote}{\fnsymbol{footnote}}
\begin{LARGE}
\bfseries{\sffamily \papertitle}
\end{LARGE}

\noindent \rule{\textwidth}{.6mm}

\vspace*{1.6cm}

\noindent \begin{large}%
\textsf{\bfseries%
\paperauthor
}
\end{large}


\phantom{XX}
\begin{minipage}{.8\textwidth}
\begin{it}
\noindent Instituto de Ciencias Nucleares \\
Universidad Nacional Aut\'onoma de M\'exico\\
Apdo. Postal 70-543, 04510 M\'exico, D.F., MEXICO \\
\end{it}
\texttt{chryss@nuclecu.unam.mx, sudarsky@nuclecu.unam.mx
\phantom{X}}
\end{minipage}
\\

\vspace*{3.5cm}
\noindent
\textsc{\large Abstract: }
The issue of whether  some manifestations of gravitation in the
quantum domain, are indicative
or not of a non-geometrical aspect in gravitation is discussed. We
examine some examples that have
been considered in this context, providing a critical analysis of
previous interpretations. The
analysis of these examples is illustrative about certain
details in the
interpretation of quantum mechanics.  We conclude that there are, at
this time, no indications of
such departure from the geometrical character of gravitation.
\end{titlepage}
\setcounter{footnote}{0}
\renewcommand{\thefootnote}{\arabic{footnote}}
\setcounter{page}{2}
\noindent \rule{\textwidth}{.5mm}

\tableofcontents

\noindent \rule{\textwidth}{.5mm}
\section{Introduction}
\label{Intro}
\small \normalsize Gravitation is the only one of the known forces for 
which there is at present no fully
satisfactory quantum description, despite the recent progress that
some of
the approaches to the
problem have achieved.  At some point, workers in the field  even 
wondered whether gravity should be quantized at
all or whether it should,  instead, be considered as an effective
phenomenon, of the type described, for example, by
thermodynamics \cite{Jac:95} ---  in this latter case it would  be
clearly inappropriate to attempt a
description at the quantum level. Fundamental considerations naturally
enter in the analysis of such questions. For instance,
if gravity could be thought to be classical in all situations,
then one could use the gravitational field associated with a particle
to determine its location
and state of motion to an accuracy higher that that allowed by the
uncertainty principle. Another line of attack involves
experiments that probe some aspects of gravitation at the
quantum level. Although recent ideas are suggestive of possible
experimental indications of
quantum gravitational  features \cite{Ame:00},  the  aspect that has
in fact been  subject to
experiment is the opposite end of the problem, namely, the effect of
gravitation on the behavior of
quantum systems. The most famous among these tests is the COW
\cite{Col.Ove.Wer:75} experiment, in which a
gravity-induced quantum phase difference is measured by the
interference of two neutron beams
that travel through regions of different gravitational potential.
This remarkable confirmation of some of the basic assumptions inherent
to quantum mechanics and to  gravitation led to a whole
series
of other related experiments, either real (see,
\eg{}~\cite{Nes.Boe.Pet:02} and
references therein)
or {\it gedanken}~\cite{Alh:97}. The hope behind all these efforts
is to draw lessons about the behavior of gravity in the quantum
realm.

One issue that takes center stage in these discussions is whether
gravity maintains its geometrical nature
at the quantum level. For instance, a well-known quantum
mechanics textbook~\cite{Sak:94} asserts that the outcome
of the COW experiment is indicative of a non-geometric aspect of
gravitation, since the effect measured is sensitive to the mass of the
particles used.  A more puzzling case is the {\em gedanken}
use of neutrino oscillations (between different mass
eigenstates) to construct
clocks that do not ``red-shift'' in a universal way under the
influence of gravitational fields~\cite{Alh:97}.
The aim of this paper is to  provide a well defined set of criteria in
order to determine whether particular
gravitational effects can be said to be of a non-geometrical nature,
and to
examine in this light the
situations that have been thought to reflect such  non-geometrical
aspects.
\section{Geometry and Gravitation}
\label{GeoG}
The basic idea behind the geometrical  description of gravity is the
equivalence principle (E.P.),
which leads to the possibility of constructing effective locally
inertial frames (LIFs) about  every point
in space-time. In practice, the construction itself of LIFs is tied to
the EP, given the fact that it is impossible to isolate any particle
from the influence of gravity and thus to obtain any object that
can be considered free of
interactions, for which the law of inertia might have meaning. 
Therefore, the EP is
not to be regarded as a circumstantial feature of  the gravitational
interaction, but as part of
the foundations of physics itself. 

The geometrical character of gravitation then is tied with the
possibility of simulating
gravitation by using  an accelerated frame. Equivalently, one can
``cancel" gravitation by accelerating without constraints under the
influence of gravity, \ie, by going to a freely falling frame%
\footnote{%
Notice that going to a ``freely falling frame''
involves, in general, an appropriate state of rotation, in addition to an
appropriate state of acceleration.%
}.
The experiments mentioned in the introduction measure, in one way or
another, quantum effects in the presence of a gravitational field. In
view of the above discussion, our general strategy will be to look at
these experiments from a LIF and compare the results with the case of
zero gravitational field.

Let us start by being more precise with our notion of geometricity
of gravitation. We must clarify that we seek a
phenomenological definition, appropriate for examining both
real and {\em gedanken} experiments, and not a theoretical
one, such as the one given in terms of the requirement of general
covariance~\cite{Wal:84}. The latter, although both appropriate
and useful when considering a given
classical theory, would be neither when applied to the analysis of
certain experimental situations, in particular those pertaining to the 
quantum domain,
\eg, in the possible
quantum gravity induced modifications of the dispersion relations of
photons~\cite{Ame:00}.

One can start by saying simply that

\begin{definition}{D1}
Gravity is geometrical if all its effects can be locally
canceled  (or simulated) by a suitable  choice of the
reference frame in which their description takes place.
\end{definition}

\noindent  One needs to be careful with the above statement.
For one, 
care is needed in treating the notion of
locality, even at the classical level, when dealing, for instance, with
particles interacting with electromagnetic fields.  Second, one can
imagine experimental devices designed to measure  some components of the
Riemman tensor, \eg, a differential accelerometer, the behavior of
which in a curved spacetime  could never be simulated
in a flat one. Third, one should take into account the possibility that
there might exist new kinds of matter%
\footnote{%
We have in mind particles associated with supersymmetry
or matter described by the cosmological fields (quintessence and
the like) that have been considered as candidates for
the ``dark energy".%
} 
with nonminimal coupling to gravitation, which, just like a 
differential accelerometer, would behave differently in a flat than in 
a curved
spacetime, but the existence of which would in no way put in question 
the geometrical character of
gravitation. Finally, one should note that the above definition, 
natural as it may be in a classical context, can in
principle become suspicious when applied to quantum systems, because of the  
quantum limits on localizability of the latter. 

In view of the above discussion,
a general formal definition of  what is to be understood by the
geometricity of gravitation will not be attemted here. Rather, we
focus on a set of useful criteria, that permit the analysis of the
experiments mentioned in the introduction.
As a first step, we concentrate  on the propagation of a ``free particle",
the latter being defined as any particle (including, \eg, atoms and 
subatomic particles)  with no
electric charge (or higher multipole moments), 
separated from other particles by distances that ensure that the nuclear
forces are not operative. We propose the following definition


\begin{definition}{D2}  
Gravitation is geometrical in nature, if the
description of the propagation of free particles given by their wave
function, in a given gravitational field, is identical to the one 
obtained in an appropriately moving frame in Minkowski spacetime.
\end{definition}

Some qualifications are necessary. First, the description 
mentioned above should refer to a region of spacetime which is large
compared with that over which the particle's wavefunction is
appreciable. At the same time, it should be 
small compared with the region over which tidal effects become
appreciable. Finally, any discrepancies should scale in an appropriate
way (which depends on the actual quantity being measured) with the
size of the region in question. In fact, with such corrections it
should be possible 
to determine the Riemann tensor in the above region. 

We can now deal with more complex systems, 
such as devices sensitive to tidal forces, or
matter with nonminimal couplings. In those cases the geometrical
character of gravitation will be tied to the notion that all of the local
effects of gravitation should be accounted for by the Riemann tensor, and in
particular, that if the Riemann tensor vanishes in a region, all
experiments carried out completely within that region should be exactly
reproducible in an appropriately moving frame in flat spacetime. 

As an
example, we apply the above concepts to the following statement

\begin{statement}{S1}
The geometrical nature of gravity requires that all
clocks red-shift in a universal way under its influence.
\end{statement}

\noindent Imagine a (poorly designed) clock, sensitive to second 
derivatives of the
gravitational potential (\ie, tidal forces). Such a clock does 
not red-shift in the same way as, say, an atomic clock and, therefore,
(S\ref{S1}) would lead us to conclude that gravity in not geometrical
in nature. On the other hand, our criterion (D\ref{D2}) provides for
discrepancies between the two clocks that scale appropriately with the
size of the clocks. Clearly, the difference in red-shifts between the
two clocks falls within this provision, and therefore, (S\ref{S1}) is
incompatible with (D\ref{D2}). We will say accordingly that
(S\ref{S1}) is incorrect.

Having said all this we proceed to examine a number of experiments
which, it has been argued, indicate non-geometrical aspects of
gravitation. We
will show that these interpretations are not appropriate, in the sense
advocated earlier.
Actually, given the fact that the experiments do not
involve tidal
effects, it is enough to consider them in light of 
(D\ref{D1}), even
though
in principle, and given the quantum nature of the probes, one should
rely on the more refined definition (D\ref{D2}).
We will refer, in the discussion of the last section,  to a new class
of 
possible experiments, the analysis of which would rely in an essential
way on (D\ref{D2}).

As a final point, we would like to stress that even though one is, of
course, free to use a different definition of ``the geometrical 
character of gravitation",
we believe that the notion expressed by (D\ref{D2}) is the closest 
in spirit to the general relativistic one, as applied in the realm
of classical physics, and 
is such that one can expect it to be appropriate for the quantum
domain.
\section{The COW Experiment}
\label{CE}
Next we turn to this famous neutron intereferometry experiment, in
which one considers the interference pattern of two
neutron beams that travel on two paths on a  plane.  The observational
quantities are related to change in this interference
pattern  when the plane is rotated in such a way that, at one moment
the
plane is perpendicular to the gradient of the earth's
gravitational potential,  and at some other instant it  is
tangent to
it. In this way one measures the  dependence of
the phase difference in the two neutron beams on  the gravitational
field
of the earth. The result turns out to depend on the
neutron's mass, and this has lead to  interpretations of this
experiment as
showing a non geometrical aspect of gravity.
Underlying such interpretations there is a notion that

\begin{statement}{S2}
The geometrical nature of gravitation should make it impossible
to determine the mass
of a particle through the use of purely passive gravitational effects.
\end{statement}

\noindent Here the
problem is due to the failure to recognize that, at the
quantum level, the mass of a particle is associated with a geometrical
scale. In fact we can determine, even in the absence of gravity, the
mass of a particle
by purely geometrical means, \ie, relying only on the
behavior of freely propagating
particles and not on properties of their interactions: 
take a
monochromatic particle beam and measure its momentum
$p$ by a simple double slit experiment. Then make the double slit
experimental
set-up move with velocity $V$ with respect to the
laboratory, and measure the momentum $p'$ of the beam in the new
frame --- the mass is read off as $M =\frac{p'-p}{V}$.
Let us apply our criterion (D\ref{D2}): the dimensions of the
apparatus used are much smaler than the scale over which it could
detect tidal effects, 
and the neutron wavelength is much smaller than that.
Thus, we should be able to account for the  experimental
result from the point of view of an inertial observer who
watches the entire apparatus moving upwards with constant acceleration $g$.
The authors of \cite{Col.Ove.Wer:75} allude to such a description ---
we give a short outline of ours for completeness.

Referring to Fig.~\ref{COWfig}, suppose that the apparatus is
accelerating upwards and has, momentarily, zero velocity. Then the
wave vector of the beam at the lower part of the segment $AB$, 
right above $A$, is equal
to the incoming value $k_0$. Further up that segment, for a fixed time
$t$, the wave vector decreases linearly with the height because of
Doppler shift (when it was emitted by the beam splitter at $A$, the
latter was moving downwards). The wave on the horizontal segment $BD$,
immediately to the right of $B$, suffered no Doppler shift when it was
reflected at $B$ because, by assumption, $B$ is momentarily at rest.
One might think that at points further to the right on $BD$ the
wave vector
will keep decreasing but this is not so: when the wave at, say, point
$P$ was reflected by the mirror at  $B$, the latter was moving
downwards with a velocity that is bigger the further to the right $P$
is. A simple calculation shows that the Doppler shift suffered in this
second reflection cancels the $x$-dependence of the wave vector.
Similar remarks apply to the $ACD$ path and the calculation of the
resulting phase difference at $D$ is algebraically identical with the
one in the presence of gravity.
\setlength{\figurewidth}{.5\textwidth}
\begin{figure}
\centerline{%
\begin{pspicture}(-.3\figurewidth,-0.1\figurewidth)%
                 (.9\figurewidth,.61\figurewidth)
\setlength{\unitlength}{.25\figurewidth}
\psset{xunit=.25\figurewidth,yunit=.25\figurewidth,arrowsize=1.5pt 3}
\psline[linewidth=.2mm]{->}%
(0,0)(0,2.4)
\put(-.1,2.4){\makebox[0cm][r]{$y$}}
\psline[linewidth=.2mm]{->}%
(0,0)(2.9,0)
\put(2.95,-.1){\makebox[0cm][l]{$x$}}
\psline[linewidth=2mm,linecolor=lightgray]{-}%
(-.2,-.2)(.2,.2)
\psline[linewidth=.5mm]{-}%
(0,0)(2,0)
\psline[linewidth=.5mm]{->}%
(.9,0)(1.05,0)
\psline[linewidth=.5mm]{-}%
(2,0)(2,1.5)
\psline[linewidth=.5mm]{->}%
(2,.6)(2,.8)
\psline[linewidth=.5mm]{-}%
(2,1.5)(0,1.5)
\psline[linewidth=.5mm]{->}%
(.4,1.5)(.6,1.5)
\psline[linewidth=.5mm]{->}%
(1.4,1.5)(1.6,1.5)
\psline[linewidth=.5mm]{-}%
(0,1.5)(0,0)
\psline[linewidth=.5mm]{->}%
(0,.6)(0,.8)
\pscustom[fillstyle=solid,fillcolor=gray]{%
\psline[linewidth=.2mm]{-}%
(1.81,-.21)(2.21,.19)(2.3,.1)(1.9,-.3)(1.81,-.21)
}
\pscustom[fillstyle=solid,fillcolor=gray]{%
\psline[linewidth=.2mm]{-}%
(-.21,1.31)(.19,1.71)(.1,1.8)(-.3,1.4)(-.21,1.31)
}
\psline[linewidth=.5mm]{-}%
(-1,0)(0,0)
\psline[linewidth=.5mm]{->}%
(-.7,0)(-.5,0)
\rput(-.15,.03){\rnode{AA}{}}
\rput(-.03,0.15){\rnode{BB}{}}
\rput(-.4,.4){\rnode{CC}{}}
\nccurve[angleA=-50,angleB=120,linewidth=.18mm]{->}{CC}{AA}
\nccurve[angleA=-45,angleB=165,linewidth=.18mm]{->}{CC}{BB}
\put(-.42,.42){\makebox[0cm][r]{$k_0$}}
\put(.06,-.22){\makebox[0cm][l]{$A$}}
\put(2.14,-.22){\makebox[0cm][l]{$C$}}
\put(2.1,1.5){\makebox[0cm][l]{$D$}}
\put(-.15,1.63){\makebox[0cm][r]{$B$}}
\psdots[dotsize=3pt]%
(1.2,1.5)
\put(1.2,1.55){\makebox[0cm][l]{$P$}}
\psline[linewidth=.5mm]{->}%
(1,1.5)(1,2)
\put(0.95,2){\makebox[0cm][r]{$g$}}
\rput(.03,1.35){\rnode{A}{}}
\rput(.15,1.47){\rnode{B}{}}
\rput(.4,1.1){\rnode{C}{}}
\nccurve[angleA=140,angleB=-30,linewidth=.18mm]{->}{C}{A}
\nccurve[angleA=135,angleB=-75,linewidth=.18mm]{->}{C}{B}
\put(.43,1){\makebox[0cm][l]{$k_1$}}
\end{pspicture}%
}
\capitem{%
The COW experiment, as seen from an inertial frame. The wave at $B$
was split at $A$ when $A$ was moving downwards with velocity $g
a/v_0$. The wave vector is Doppler red-shifted, $k_1=k_0 (1-\epsilon)$,
$\epsilon \equiv g
a/v_0^2$. The transition from $k_0$ at $A$ to $k_1$ at $B$ is linear
in the height
$y$. The wave at $P$ was reflected at $B$ when $B$ was moving
downwards with velocity $g x_P/v_1 = g x_P/v_0 + \calO
(\epsilon^2)$. The resulting Doppler blue-shift cancels the
$x$-dependent part of the red-shift at $A$ and makes the
wave vector constant along $BD$. The contributions to the phase from
$AB$ and $CD$ cancel, so that $\Delta \phi = (k_0 - k_1)b = \epsilon
k_0 b$.
}
\label{COWfig}
\end{figure}
Thus, according to (D\ref{D2}), the COW experiment supports rather than
negates the geometrical nature of gravitation and (S\ref{S2}) is
therefore incorrect.
\section{The Neutrino Oscillation Clocks}
\label{NIC}
Recently,  a {\em gedanken } experiment has been considered, in which
two clocks that base their
operation on  neutrino oscillations ``red-shift'' differently due to
the effects of gravitation~\cite{Alh:97}. This has been interpreted as an
indication of a non-geometrical aspect of gravitation. We will examine
this assertion using the criterion set forth in~(D\ref{D2}).

The set-up uses the Lense-Thirring effect --- we avoid inessential
complications by considering a simplified setting in which this is the
only relevant gravitational effect present.
Consider a hollow
spherical shell with mass $M$ and radius $R$ ($R>>MG$), rotating about the
$z$-axis with angular velocity $\vec{\alpha}$.
We have, besides the constant gravitational potential
$\Phi_0= -GM/R $ (relative
to points at infinity), a gravitomagnetic field $\vec B$, given by
$\vec B= \frac{2 GM}{3R} \vec\alpha$.
The spacetime metric inside the rotating shell,
in first order perturbation theory (with both $\Phi_0$ and $B$ of
first order), is
\ble{metric}
dS^2 = \left[- (1-2\Phi_0)dt^2 + dx^2 + dy^2 + dz^2 \right]
 + 2 B ( y dx -x dy ) dt
\, ,
\ee
where $B = |\vec B|$. One now constructs two clocks, $I$ and $II$,
that base their operation on oscillations
between suitable superpositions of the mass {\em and} spin
eigenstates $|m_i, \, \hat{z} \, \pm>$  ($m_1$,  $m_2$ are the
mass eigenvalues and $\hat{z} \, \pm$ refer to the $S_z$ eigenstates
--- we will assume $m_1 > m_2$).
Clock $I$ oscillates between the states%
\footnote{%
We have set, for simplicity, the ``mixing angle'' $\theta$
of~\cite{Alh:97} equal
to $\pi/4$ --- the results do not depend essentially  on this choice. %
}
\bae
\ket{Q_a}
& \equiv &
\sqrti \ket{m_1, \, \hat{z} \, +}
+ \sqrti \ket{m_2, \, \hat{z} \, +}
\ff
\ket{Q_b}
& \equiv &
- \sqrti \ket{m_1, \, \hat{z} \, +}
+ \sqrti \ket{m_2, \, \hat{z} \, +}
\, ,
\label{statesI}
\eae
while clock $II$ uses the pair
\bae
\ket{Q_A}
& \equiv &
\sqrti \ket{m_1, \, \hat{z} \, +}
+ \sqrti \ket{m_2, \, \hat{z} \, -}
\ff
\ket{Q_B}
& \equiv &
- \sqrti \ket{m_1, \, \hat{z} \, +}
+ \sqrti \ket{m_2, \, \hat{z} \, -}
\, .
\label{statesII}
\eae
In other words, in the four-dimensional state space available, with
basis $\{\ket{m_1, \, \hat{z}+}, \, \ket{m_1, \, \hat{z}-}, \,
\ket{m_2, \, \hat{z}+}$, $\ket{m_2, \, \hat{z}-}\}$, each clock works
in a two-dimensional
subspace. That the time evolution of the clocks does not lead outside
of this subspace will be obvious, since the states appearing in the
\rhs{} of~(\ref{statesI}), (\ref{statesII}), will be eigenstates of all
hamiltonians considered in the sequel.
Clock $I$ runs by monitoring the transition
$a \to b$, which occurs with probability $P(a \to b)$. For example, 
an ensemble of particles oscillating
between $\ket{Q_a}$ and $\ket{Q_b}$ could be observed,
with  clock $I$ ``ticking'' every time a maximum in the population of
the state $\ket{Q_a}$ is observed --- similar remarks apply to clock
$II$.

In the absence of any gravitational field the
time-evolution of the clocks is described by the Hamiltonian
\ble{H1}
{\bf H}_0 = {\bf m} c^2
\, ,
\ee
where ${\bf m}$ is the mass operator and we neglect the kinetic
term.
In this case, the spin degree of freedom is ``spectator'' and the two
clocks tick with the same frequency $2\omega_0 = \omega_{I} =
\omega_{II}$, where%
\footnote{%
One gets $P_{a \to b} = \sin^2 \omega_0 t = (1-\cos 2\omega_0 t)/2$ =
$P_{A \to B}$.%
}
\ble{omzero}
\omega_0 = \frac{(m_1 -m_2) c^2}{2 \hbar}
\equiv \frac{\Delta m c^2}{2 \hbar}
\, .
\ee
Now we place the clocks
inside the rotating shell, where the  Hamiltonian is
\ble{H2}
{\bf H}' ={\bf m}c^2 (1-2\Phi_0) + \vec {\bf S} \cdot \vec B
\ee
($\vec {\bf S}$ is the particle spin operator).  For clock $I$ the spin is
still ``spectator'' and its frequency is simply multiplied by the
factor $\lambda \equiv 1-2\Phi_0$,
\ble{omonep}
\omega_I' = \lambda \omega_I
\, .
\ee
Clock $II$, on the other hand, receives an additional shift in its
frequency by the energy difference of the two spin eigenstates that
enter in $\ket{Q_A}$, $\ket{Q_B}$,
\ble{omtwop}
\omega_{II}' = \lambda \omega_{II} + B
\, .
\ee
Does this effect indicate a non-geometric aspect of gravity? Let us apply
again the criterion (D\ref{D2}). To start with, the particle
system is assigned a length scale  of order $1/m$, which we consider
fixed once and for all.  The gravitational field
has a length scale given by $R$, but the gravitomagnetic field
inside the shell can be kept constant while scaling both $M$ and $R$
simultaneously upward until $R>>1/m$. Thus the effect remains unchanged
when we  arrange the
scales to satisfy the requirements of (D\ref{D2}).
The issue is then,
does the effect persist when one moves to an inertial reference frame,
the latter being defined as one where the metric becomes locally
Lorentzian? 

Going to  a (primed) frame rotating with an arbitrary angular velocity
$\Omega\hat{z}$,
\ble{rotfr}
t=(1+\Phi_0) t'
\, ,
\qquad
x= x' \cos \Omega t'- y'  \sin \Omega t'
\, ,
\qquad
y=y' \cos \Omega t'+ x' \sin \Omega t'
\, ,
\quad
z=z'
\, ,
\ee
the metric of (\ref{metric}) becomes
\be
dS_{\text{rot}}^2 =
-\big(1+ (2 B \Omega - \Omega^2)(x'^2 + y'^2)\big) dt'^2
+ dx'^2 + dy'^2 + dz'^2
+ 2 (\Omega -B) (x' dy' - y' dx') dt'
\, .
\ee
Choosing therefore $\Omega = B$, we find ourselves in a LIF
(in the vicinity of the origin).
If the two clocks then are rotated as above, would the equality of
their ticking rates be restored?
At first sight, one might argue as follows: First, as the eigenstates of the
hamiltonian~(\ref{H2}) 
have their spins along the $z$ axis, they should not be  
affected by the rotation of the
frame from which we now describe them. Second, the two
notions of time
(associated with the two, relatively rotating observers) coincide
on this axis. Therefore,  one might conclude that the effect would
persist 
in the freely falling frame. This
would be  very
puzzling to say the least. However, we must be careful and note that if
all we do
is change the frame of description, but not make the experimental
apparatus
(including the detectors) move with the locally inertial frame, then
the above
mentioned situation would ensue. On the other hand, if we make the
experimental apparatus
(in particular, the detectors)  move together with the locally inertial
frame, then the
effect will indeed disappear as it should.
Right from the outset, we can see that it is not unreasonable,
{\em a priori}, to expect
this, because, in the rotating primed frame,
the $S_z$-eigenstates  are described as 
\be
\ket{\hat{z} +} \to e^{i\Omega t/2} \ket{\hat{z}'+}
\, ,
\qquad
\qquad
\ket{\hat{z}-} \to e^{-i\Omega t/2} \ket{\hat{z}'-}
\, . 
\ee
The  point is that
{\em
the description
of the time evolution of a given state, is different in the rotating and
nonrotating frames, despite the fact that, on the $z$ axis, where
the particles can be thought to be located
for all practical purposes,
the two notions of time coincide%
}. As $\ket{Q_a}$ and
$\ket{Q_b}$ involve only a single spin eigenstate, the phase factor
introduced by a
rotation has no observable effect and, therefore, clock $I$ should be
insensitive to rotations. On the other hand, $\ket{Q_A}$, $\ket{Q_B}$
involve both spin eigenstates, each of which transforms with a
different phase factor, so that clock $II$ should be, in principle,
affected by rotations.

All this becomes clear if
we  give a more detailed description of how exactly are the two
clocks supposed to operate. Mass oscillations are due to the fact that
the experimentally observed particle ``flavors'', which we denote by
$\ket{\partp}$ and $\ket{\partm}$, are linear combinations of
the mass eigenstates $\ket{m_i}$,
\ble{plusminus}
\ket{\partp} = \sqrti \ket{m_1} + \sqrti \ket{m_2}
\, ,
\qquad \qquad
\ket{\partm} = -\sqrti \ket{m_1} + \sqrti \ket{m_2}
\, .
\ee
Written in terms of these, the states of clock $I$ become
\bae
\ket{Q_a}
& = &
\ket{\partp, \,  \hat{z} \, +}
\ff
\ket{Q_b}
& = &
\ket{\partm, \, \hat{z} \, +}
\, ,
\label{statesIp}
\eae
\ie, they  correspond to two different flavors with
spin along $\hat{z}$. The interesting news is that the states of
clock $II$ become
\bae
\ket{Q_A}
& = &
\sqrti \ket{\partp, \, \hat{x} \, +}
+ \sqrti \ket{\partm, \, \hat{x} \, -}
\ff
\ket{Q_B}
& = &
 \sqrti \ket{\partm, \, \hat{x} \, +}
+ \sqrti \ket{\partp, \, \hat{x} \, -}
\, ,
\label{statesIIp}
\eae
\ie, they now involve spins pointing along $\hat{x} \, $! (see
Fig.~\ref{QA}).
\setlength{\figurewidth}{.6\textwidth}
\begin{figure}
\centerline{%
\begin{pspicture}(-.1\figurewidth,-0.1\figurewidth)%
                 (.9\figurewidth,.45\figurewidth)
\setlength{\unitlength}{.25\figurewidth}
\psset{xunit=.25\figurewidth,yunit=.25\figurewidth,arrowsize=1.5pt 3}
\put(-.5,.5){\makebox[0cm][r]{$\ket{Q_A} \, \, \, = $}}
\psline[linewidth=.2mm]{->}%
(0.5,0.5)(1.5,0.5)
\put(1.45,.35){\makebox[0cm][l]{$y$}}
\psline[linewidth=.2mm]{->}%
(0.5,0.5)(0.5,1.5)
\put(.4,1.4){\makebox[0cm][r]{$z$}}
\psline[linewidth=.2mm]{->}%
(0.5,0.5)(0,0)
\put(-.1,-.05){\makebox[0cm][r]{$x$}}
\psline[linewidth=.5mm]{->}%
(0.5,0.5)(.5,1.1)
\psline[linewidth=.5mm]{->}%
(0.5,0.5)(.5,-.1)
\put(.65,1){\makebox[0cm][l]{$\ket{m_1}$}}
\put(.65,-.05){\makebox[0cm][l]{$\ket{m_2}$}}
\put(1.9,.5){\makebox[0cm][l]{$=$}}
\psline[linewidth=.2mm]{->}%
(3,0.5)(4,0.5)
\put(3.95,.35){\makebox[0cm][l]{$y$}}
\psline[linewidth=.2mm]{->}%
(3,0.5)(3,1.5)
\put(2.9,1.4){\makebox[0cm][r]{$z$}}
\psline[linewidth=.2mm]{->}%
(3,0.5)(2.5,0)
\put(2.4,-.05){\makebox[0cm][r]{$x$}}
\psline[linewidth=.5mm]{->}%
(3,0.5)(2.7,.2)
\psline[linewidth=.5mm]{->}%
(3,0.5)(3.3,.8)
\put(2.8,.05){\makebox[0cm][l]{$\ket{\partp}$}}
\put(3.35,.8){\makebox[0cm][l]{$\ket{\partm}$}}
\end{pspicture}%
}
\capitem{Equivalent ways of describing the state $\ket{Q_A}$. In each
case, a sum over the states shown is implied (see
Eqs.~(\ref{statesII}), (\ref{statesIIp})). A similar figure can be drawn
for $\ket{Q_B}$.%
}
\label{QA}
\end{figure}
This is the
fundamental difference between the two clocks, namely, for clock $II$,
a change in the
basis in the mass space affects the direction of the spin as well
(which is not true for clock $I$). The effect can of course be
traced to the fact that while the states of clock $I$ factorize in the
two spaces (mass and spin), those of clock $II$ do not, but rather,
involve sums over factorizable states.

Returning to the clock operation, we may now further specify that
clock $I$ sends a beam of particles, travelling along the $z$-axis,
towards a detector of ``top'' particles and ticks whenever a maximum
counting rate is reached.
Clock $II$ does the same,
but first passes the beam through a Stern-Gerlach
apparatus which filters out the $\ket{ \hat{x} \, -}$ component%
\footnote{%
Notice that the state
$\ket{\partp, \, \hat{x} +}$, which is the one detected by the above
procedure, only enters in $\ket{Q_A}$
(see~(\ref{statesIIp})), so we can use
it as a ``tag'' for $\ket{Q_A}$ (see also Eq.~(\ref{tpQAt})).%
}.
It is clear from this description that clock $I$ is not affected by
rotations around the
$z$-axis, as we expected. For clock $II$, we note that when rotated
with angular velocity $\Omega \hat{z}$, the (rotating) Stern-Gerlach
apparatus will block the $\ket{\hat{n} \, -}$-component, where
$\hat{n}=(\cos \phi, \, \sin \phi, \, 0)$ and $\phi=\Omega t$.
It becomes obvious then that, when rotated as above, clock $II$ actually
detects oscillations between the {\em time dependent} states
$\ket{\tilde{Q}_A}$, $\ket{\tilde{Q}_B}$, given by
\bae
\ket{\tilde{Q}_A}
& = &
\sqrti \ket{\partp, \, \hat{n} \, +}
+ \sqrti \ket{\partm, \, \hat{n} \, -}
\ff
\ket{\tilde{Q}_B}
& = &
 \sqrti \ket{\partm, \, \hat{n} \, +}
+ \sqrti \ket{\partp, \, \hat{n} \, -}
\, ,
\label{statesIIpp}
\eae
or, in terms of $\ket{Q_A}$, $\ket{Q_B}$,
\bae
\ket{\tilde{Q}_A}
& = &
\cos \frac{\phi}{2} \ket{Q_A} + i \sin \frac{\phi}{2}  \ket{Q_B}
\ff
\ket{\tilde{Q}_B}
& = &
i \sin \frac{\phi}{2} \ket{Q_A} + \cos \frac{\phi}{2}  \ket{Q_B}
\, .
\label{statesIIppp}
\eae
One  easily shows that%
\footnote{%
$\ket{Q_A, \, t}$ is the time-evolved ket that, at $t=0$, coincides
with  $\ket{Q_A}$.%
}
\ble{tpQAt}
\ipq{\tilde{Q}_A}{Q_A, \, t}
\, = \,
2 \ipq{\partp, \, \hat{n} \, +}{Q_A, \, t}
\, = \,
e^{-i \lambda c^2 (m_1 + m_2)t/(2 \hbar)}
\cos (\frac{\omega'_{II} - \Omega}{2} t)
\, ,
\ee
so that the frequency $\Omega$ of the rotation adds to
the ticking frequency $\omega_{II}'$ of clock $II$.
When the latter is
rotated with $\Omega=B$, the above rotation-induced shift in
its frequency exactly cancels the effect of the gravitomagnetic
field $\vec B$ and the two clocks tick synchronously again.
Thus, the effect described in this section is purely geometrical according
to the notion of geometricity proposed in (D\ref{D2}).
\section{Discussion}
\label{D}
Before concluding we would like to return to a point briefly
mentioned in the introduction:
the fact that the equivalence principle is at the foundations of
mechanics. We recall that
the starting point for the construction of the edifice of
classical physics is Newton's three laws,
which hold in an inertial frame. How are we supposed,
in practice, to find such a
frame?. The ``frame of the fixed stars'', that was considered 
in  Newton's time,
would clearly not be an appropriate starting point nowadays. One way 
to do it is
to take three freely moving, non-colinear particles and adjust 
the motion of our frame so as to ensure that,
relative to it, the three particles move
according to the law of inertia. At first, this might seem to
reduce the law of inertia to a mere
definition. However, its content lies in its predictive power
regarding the motion of other free particles. If there was no
gravitation we could equip ourselves
with the required free particles by  choosing them electrically neutral
(including higher electric or magnetic multipole
moments),  and ensuring that they were 
sufficiently distant from other particles so that the nuclear forces
could be neglected. However, gravitation
exists, so the problem of constructing an inertial frame persists. The EP 
is what saves the situation: we simply follow the
procedure as if there was no gravitation, and the result is a LIF.
Moreover, this
is the only way to obtain a LIF, unless one such frame is already
known and a second one is obtained by
moving inertially with respect to the first. In
this way, we see that a test of, say, the
universality of free fall, using classical objects, should
be regarded, if we want to be precise, as a test of the law of inertia.

In the above sense, one is never 
observing gravitational effects in any local experiment, for any such  
experimental manifestation simply indicates the failure to construct a
LIF. 
Gravitation manifests itself only in the impossibility to  extend our
LIF to a global inertial frame,
\ie, in its tidal effects --- this is of
course nothing but the general
relativistic lesson that gravitation resides in the Riemman curvature
tensor. 
This remark applies, in particular, to the COW experiment, 
which, in our view, only confirms  that the
above  procedure to construct a LIF, where the laws of
mechanics are valid,  {\em yields at the
same time a frame in which Schroedinger's equation is valid},  
certainly a highly nontrivial result.

In conclusion, we identify the reason for the sufficiency of
(D\ref{D1}) in our analysis so far:
none of the experiments that have been carried out to date,
as far as we know, attempts to detect gravitational tidal
effects using quantum mechanical probes.
This is a serious shortcoming of our experimental knowledge in this
field, especially if we note that, in accordance
to the discussion above, it means that we have not been
testing gravitation at all! The point
here is not to be critical in any way of the heroic efforts of our
experimental colleagues, but just to
clarify what indeed has been tested and what still lies ahead. We hope
that the challenge of detecting gravitational tidal effects on quantum
systems will soon be undertaken.
\section*{Acknowledgments}
The authors would like to acknowledge partial support from CONACyT
projects 32307-E (C.{} C.), 32272-E (D.{} S.) and DGAPA-UNAM projects
IN 119792 (C.{} C.), IN 121298 (D.{} S.).

\begin{thebibliography}{1}

\bibitem{Alh:97}
D.~Ahluwalia.
\newblock On a {N}ew {N}on-{G}eometric {E}lement in {G}ravity.
\newblock {\em Gen. Rel. and Grav.}, 29:1491, 1997.

\bibitem{Ame:00}
G.~Amelino-Camelia.
\newblock Are {W}e at the {D}awn of {Q}uantum {G}ravity {P}henomenology?
\newblock {\em Lect. Notes Phys.}, 541:1, 2000.

\bibitem{Col.Ove.Wer:75}
R.~Colella, A.~W. Overhauser, and S.~Werner.
\newblock Observation of {G}ravitationally {I}nduced {Q}uantum {I}nterference.
\newblock {\em Phys. Rev. Lett.}, 34:1472, 1975.

\bibitem{Nes.Boe.Pet:02}
V.{} V.{}~Nesvizhevsky {\em et al.}
\newblock Quantum {S}tates of {N}eutrons in the {E}arth's {G}ravitational
  {F}ield.
\newblock {\em Nature}, 415:297--299, 2002.

\bibitem{Jac:95}
T.~Jacobson.
\newblock Thermodynamics of {S}pacetime: the {E}instein {E}quation of {S}tate.
\newblock {\em Phys. Rev. Lett.}, 75:1260--1263, 1995.

\bibitem{Sak:94}
J.~J. Sakurai.
\newblock {\em Modern {Q}uantum {M}echanics ({R}evised {E}dition)}.
\newblock Addison-Wesley, 1994.

\bibitem{Wal:84}
R.~M. Wald.
\newblock {\em General {R}elativity}.
\newblock University of Chicago Press, 1984.

\end{thebibliography}

\end{document}